\documentclass[amssymb,aps,pre,manuscript,secnumarabic]{revtex4}
\usepackage{amsmath}
\usepackage{amssymb}
\usepackage{graphicx}

\begin{document}

\title{Enhanced Entrainability of Genetic Oscillators by Period Mismatch}

\author{Yoshihiko Hasegawa}

\email{hasegawa@cb.k.u-tokyo.ac.jp}

\affiliation{Department of Biophysics and Biochemistry, Graduate School of Science,
The University of Tokyo, Tokyo 113-0033, Japan}

\author{Masanori Arita}

\affiliation{Department of Biophysics and Biochemistry, Graduate School of Science,
The University of Tokyo, Tokyo 113-0033, Japan}

\date{Jan 12, 2013}
\begin{abstract}
Biological oscillators coordinate individual cellular components so
that they function coherently and collectively. They are typically
composed of multiple feedback loops, and period mismatch is unavoidable
in biological implementations. We investigated the advantageous effect
of this period mismatch in terms of a synchronization response to
external stimuli. Specifically, we considered two fundamental models
of genetic circuits: smooth- and relaxation oscillators. Using phase
reduction and Floquet multipliers, we numerically analyzed their entrainability
under different coupling strengths and period ratios. We found that
a period mismatch induces better entrainment in both types of oscillator;
the enhancement occurs in the vicinity of the bifurcation on their
limit cycles. In the smooth oscillator, the optimal period ratio for
the enhancement coincides with the experimentally observed ratio,
which suggests biological exploitation of the period mismatch. Although
the origin of multiple feedback loops is often explained as a passive
mechanism to ensure robustness against perturbation, we study the
active benefits of the period mismatch, which include increasing the
efficiency of the genetic oscillators. Our findings show a qualitatively
different perspective for both the inherent advantages of multiple
loops and their essentiality. 
\end{abstract}
\maketitle

\section{Introduction}

Biochemical oscillators exist primarily at the genetic level, such
as in the interplay among mRNAs and proteins \cite{Novak:2008:BiochemOsc}.
At least one negative feedback loop (NFL) is essential for limit-cycle
oscillation, and in the minimum configuration, one NFL with time-delay
is sufficient to generate oscillation. However, many realistic oscillators
possess extra components, such as positive feedback loops (PFL) and
more than one NFLs, that are apparently redundant. Since more regulatory
components consume more resources (e.g., amino acids for synthesis
and adenosine triphosphate for operation), this would seem energetically
disadvantageous, and thus not preferable through evolution. In addition,
the coexistence of positive and negative loops may make the resulting
behavior incoherent. Still, the circadian clock of even the simplest
life form (e.g., the marine dinoflagellate \textit{Gonyaulax polyedra})
is composed of multiple oscillators whose rhythms exhibit different
periodic patterns \cite{Roenneberg:1993:TwoOsc,Morse:1994:TwoCircadian}.
Usually this observation is explained by the notion of biological
robustness, in which there is a backup mechanism or a resistance to
perturbation. However, such observations do not suggest the essentiality
of multiple loops: why does not even a single species possess the
minimum configuration? In this study, we numerically analyze the inherent
advantages of multiple loops from the perspective of entrainment and
the adaptation of the system to external input \cite{Gonze:2000:Entrainment}.

Biologists have investigated coupled oscillators at the molecular
level in many different organisms. Bell-Pedersen \emph{et al.} reviewed
circadian clocks in several species from different clades \cite{Pederse:2005:MultipleCircadian}.
They suggested that multiple loops comprise pacemaker and slave oscillators
in unicellular organisms, whereas in mammals and avians, a centralized
pacemaker entrains downstream systems. Circadian systems have been
also investigated theoretically. Wagner \emph{et al.} studied the
stability of an oscillator (called the Goodwin model \cite{Goodwin:1965:Oscillator})
against perturbation of the kinetic rate, and found that interlinked
loops are more robust (by nearly one order of magnitude) than a non-linked
counterpart~\cite{Wagner:2005:TwoGeneOsc}. Other benefits of multiple
feedback loops are found in Refs~\cite{Trane:2007:RobustCircadian,Hafner:2010:MultiLoop},
and their intuitive advantages are summarized in the review by Hastings~\cite{Hastings:2000:TwoLoopReview}.

When multiple loops are involved, synchronization becomes a crucial
issue. In biological implementations, no loops can share an identical
period in a strict sense. At first glance, such \emph{disorder} seems
to degrade the system performance. However, the results in nonlinear
physics indicate that a certain amount of disorder actually enhances
performance through mechanisms similar to those in stochastic resonance
\cite{Benzi:1981:SR,McNamara:1989:SR,Jung:1991:AmpSR,Jung:1992:GlobalSR,Gammaitoni:1998:SR,McDonnell:2008:SRBook,McDonnell:2009:SR,Hasegawa:2011:BistableSIN},
Brownian transport \cite{Astumian:1994:MolecularMotor,Astumian:1997:BrownMotor,Frey:2005:BrownianRev,Hanggi:2009:BrownianMotorsReview,Hasegawa:2012:Motor},
noise-induced synchronization \cite{Marchesoni:1996:SpatiotempSR,Nakao:2007:NISinLC,Teramae:2004:NoiseIndSync},
and disorder-induced resonance \cite{Tessone:2006:DiversityResonance}.

Our focus is the property called entrainability, or the ability of
oscillators to be synchronized with an external periodic signal. It
is directly connected to the adaptability of the species and therefore
its survival \cite{Sharma:2003:CircadianAdaptive}. Not only does
a \emph{Drosophila} with a circadian clock survive better than one
without ~\cite{Beaver:2002:FlyClock}, but also an \emph{Arabidopsis}
with better entrainment with the environment has increased photosynthesis
and consequently grows faster \cite{Dodd:2005:PlantCircadian}. Even
cyanobacteria with better entrainability have an advantage over nonsynchronizable
ones, although this advantage disappears in constant environments
\cite{Woelfle:2004:CyanoClock}.

At the molecular level, entrainability has been attributed to several
different mechanisms. Gonze \emph{et al.} studied a population of
suprachiasmatic nuclei (SCN) neurons and showed that efficient global
synchronization is achieved via damping~\cite{Gonze:2005:CircadianSync}.
Liu \emph{et al.} experimentally confirmed that robust oscillatory
behavior can be achieved by tissue-level communication between SCN
cells, even if critical clock genes have been knocked out ~\cite{Liu:2007:Cell}.
Tsai \emph{et al.} reported that the presence of PFLs allowed for
easier tuning of the period without changing the amplitude~\cite{Tony:2008:TunablePositive}.
Zhou \emph{et al.} showed that positive feedback loops assist noise-induced
synchronization through their sensitivity to external signals~\cite{Zhou:2008:SyncGeneOsc}.

Still, a fundamental question has not been answered: what is the advantage
of harboring multiple loops with mismatched periods? We investigate
the effects of mismatched periods on limit-cycle oscillations from
the viewpoint of entrainment. Specifically, we consider the minimal
fundamental model: an oscillator made of two loops, each of a different
period, that are connected. By using phase reduction and Floquet multipliers,
we analyzed the response of this system to external periodic stimuli
as a function of the coupling strength and the period ratio. Our main
finding is the large effect of mismatched periods on entrainability:
that is, better entrainment is achieved when the two oscillatory components
have different natural periods. This enhancement effect is observed
in two types of coupled oscillators: smooth and relaxation oscillators
(Figure~\ref{fig:motifs}). The enhanced entrainability seen in the
present study is similar to that reported by Komin \emph{et al.}~\cite{Komin:2010:EntrainCircadian}.
However, our mechanism is different from theirs: the enhancement in
our case is caused by the oscillators in the vicinity of the \emph{bifurcation
on the limit cycle} whereas their enhancement results from damped
oscillation (oscillator death).

\section{Methods}

We studied the effects of mismatched periods on entrainment by using
smooth and relaxation oscillators. Before explaining the details of
these models, we will first explain period scaling and a coupling
scheme.

\subsection{The base model}

An $N$-dimensional differential equation for a single oscillator
is represented by 
\begin{equation}
\frac{d\boldsymbol{v}}{dt}=\boldsymbol{f}(\boldsymbol{v}),\label{eq:LC_def}
\end{equation}
where $\boldsymbol{v}$ and $\boldsymbol{f}(\boldsymbol{v})$ are
$N$-dimensional column vectors defined by 
\[
\boldsymbol{v}=(x,y,z,\cdots)^{\top},\hspace{1em}\boldsymbol{f}(\boldsymbol{v})=(f_{1}(\boldsymbol{v}),f_{2}(\boldsymbol{v}),\cdots,f_{N}(\boldsymbol{v}))^{\top},
\]
where $\top$ denotes a transposed vector (matrix). Equation~\ref{eq:LC_def}
yields autonomous oscillation (a limit-cycle oscillation) whose period
is $T$. The period can be tuned without changing the amplitude of
the oscillation by introducing a scaling parameter $\tau$ 
\begin{equation}
\frac{d\boldsymbol{v}}{dt}=\frac{\boldsymbol{f}(\boldsymbol{v})}{\tau},\label{eq:LC_mod}
\end{equation}
where the period of equation~\ref{eq:LC_mod} is given by $\tau T$.
Other than their periods, solutions to equation~\ref{eq:LC_mod}
with different $\tau$ values will have the same properties~\cite{Gonze:2005:CircadianSync,Komin:2010:EntrainCircadian}.
We used equation~\ref{eq:LC_mod} to study the effects of mismatched
periods between two oscillatory components: 
\begin{eqnarray}
\frac{d\boldsymbol{v}_{1}}{dt} & = & \frac{\boldsymbol{f}(\boldsymbol{v}_{1})}{\tau_{1}}+\boldsymbol{C}_{1}(\boldsymbol{v}_{1},\boldsymbol{v}_{2}),\label{eq:coupled_1}\\
\frac{d\boldsymbol{v}_{2}}{dt} & = & \frac{\boldsymbol{f}(\boldsymbol{v}_{2})}{\tau_{2}}+\boldsymbol{C}_{2}(\boldsymbol{v}_{2},\boldsymbol{v}_{1}),\label{eq:coupled_2}
\end{eqnarray}
where the subscript $i$ denotes the $i$th component and $\boldsymbol{C}_{i}(\boldsymbol{v}_{i},\boldsymbol{v}_{j})$
represents a coupling term. To investigate the effects of mismatched
periods, we set 
\begin{equation}
\tau_{1}=1,\hspace{1em}\tau_{2}=R,\label{eq:R_def}
\end{equation}
where $R$ represents the ratio of the periods of the two oscillators.

\begin{figure}
\includegraphics[width=15cm]{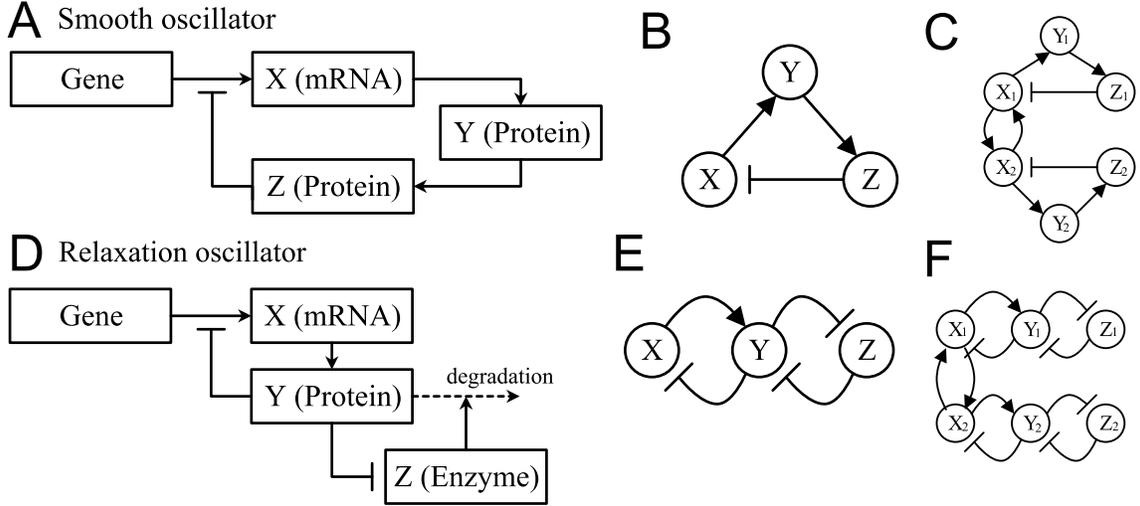}

\caption{(A) Diagram of a smooth oscillator, (B) simplified diagram of a single
smooth oscillator, (C) two coupled smooth oscillators, (D) diagram
of a relaxation oscillator, (E) simplified diagram of a single relaxation
oscillator, and (F) two coupled relaxation oscillators. The symbols
$\rightarrow$ and $\dashv$ represent positive and negative regulations,
respectively. \label{fig:motifs}}
\end{figure}

\begin{figure}
\includegraphics[width=15cm]{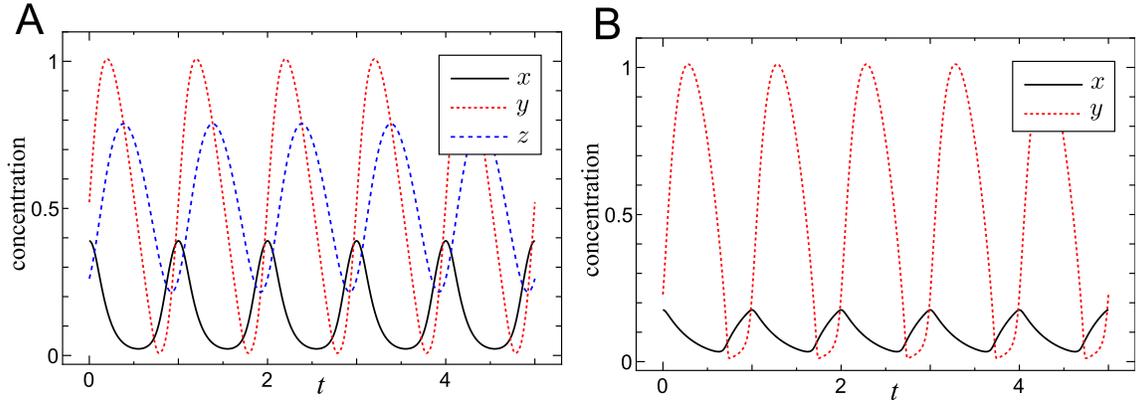}

\caption{Time courses of (A) the smooth oscillator (equation~\ref{eq:F_smooth_def})
and (B) the relaxation oscillator (equation~\ref{eq:F_relax_def}).
Solid, dotted, and dashed lines denote $x$, $y$, and $z$, respectively.
The period and the amplitude of $y$ are normalized to $1$ in both
models. \label{fig:oscillation}}
\end{figure}

\subsection{Genetic oscillator models}

We show two specific models of $\boldsymbol{f}(\boldsymbol{v})$,
each for the smooth and relaxation oscillators.

\subsubsection{Smooth oscillator}

A \emph{smooth oscillator }is the most basic structure for limit-cycle
oscillations \cite{Novak:2008:BiochemOsc}, and many biochemical oscillators
(e.g., the Goodwin model \cite{Goodwin:1965:Oscillator} and the repressilator
\cite{Elowitz:2000:Repressilator}) belong to this class. We adopted
the smooth oscillator presented in Nov\'ak and Tyson~\cite{Novak:2008:BiochemOsc}
(Figure~\ref{fig:motifs}). A diagram and the simplified structure
of the smooth oscillator are shown in Figures~\ref{fig:motifs}A
and B, where the transcription of $X$ (mRNA) is inhibited by $Z$
(protein). Protein $Y$ plays a role in the time-delay of the loop
$X\rightarrow Y\rightarrow Z\dashv X$, where $X\rightarrow Y$ and
$X\dashv Y$ denote activation and repression, respectively. In the
presence of sufficient time delay in the NFL, the regulatory network
exhibits limit-cycle oscillations. The dynamics of the smooth oscillator
are described by the following three-dimensional differential equation
on $\boldsymbol{v}=(x,y,z)^{\top}$, where $x$, $y$, and $z$ denote
the concentrations of mRNA $X$, protein $Y$, and protein $Z$, respectively:
\begin{equation}
\boldsymbol{f}(\boldsymbol{v})=\left(\begin{array}{c}
{\displaystyle V_{sx}\frac{K_{d}^{p}}{K_{d}^{p}+z^{p}}-k_{dx}x}\\
{\displaystyle k_{sy}x-V_{dy}\frac{y}{K_{m}+y}}\\
k_{sz}y-k_{dz}z
\end{array}\right).\label{eq:F_smooth_def}
\end{equation}
Because the amplitude and the period of the limit cycle can be tuned
separately by proper scaling, we employed the parameters $V_{sx}=5.44$,
$k_{dx}=8.99$, $k_{sz}=4.49$, $k_{dz}=4.49$, $k_{sy}=18.0$, $V_{dy}=2.72$,
$K_{m}=0.00303$, $K_{d}=0.303$, and $p=4.0$ to achieve unit period
and unit amplitude for $y$ (we define the amplitude as the difference
between the maximum and the minimum of the oscillation. See Figure~\ref{fig:oscillation}A).
These parameter settings are essentially identical to those presented
in the Supplement of Nov\'ak and Tyson~\cite{Novak:2008:BiochemOsc},
except for the scale transform ($V_{sx}=0.2$, $k_{dx}=0.1$, $k_{sz}=0.05$,
$k_{dz}=0.05$, $k_{sy}=0.2$, $V_{dy}=0.1$, $K_{m}=0.01$, $K_{d}=1.0$,
and $p=4.0$ in the original model).

We analyzed two coupled smooth oscillators, as shown in Figure~\ref{fig:motifs}C,
where the subscript $i$ of $X_{i}$ denotes the $i$th oscillator
component. We assumed that the protein translated from $X_{1}$ positively
regulates the transcription of $X_{2}$. The coupling term in equations~\ref{eq:coupled_1}
and \ref{eq:coupled_2} can be written in the following linear relation:
\begin{equation}
\boldsymbol{C}_{i}(\boldsymbol{v}_{i},\boldsymbol{v}_{j})=\left(\begin{array}{c}
\epsilon(x_{j}-x_{i})\\
0\\
0
\end{array}\right)\hspace{1em}((i,j)=(1,2)\,\,\mathrm{or}\,\,(2,1)),\label{eq:smooth_coupling_term}
\end{equation}
where $\epsilon$ denotes the coupling strength.

\subsubsection{Relaxation oscillator}

The existence of PFLs makes oscillators excitatory, and such oscillators
are known as \emph{relaxation oscillators} \cite{Novak:2008:BiochemOsc}.
Relaxation oscillators have cusps in their nullcline curves, and one
example of this class is the FitzHugh--Nagumo oscillator \cite{FitzHugh:1961:FHmodel,Nagumo:1962:NagumoModel}.

We again employ a relaxation-type genetic oscillator presented by
Nov\'ak and Tyson~\cite{Novak:2008:BiochemOsc}. Figures~\ref{fig:motifs}D
and E show specific and simplified models, respectively, of a relaxation
oscillator in which transcription of $X$ (mRNA) is inhibited by $Y$
(protein), which serves as a NFL. Following this, $Y$ (mRNA) is degraded
by $Z$ (enzyme) and also forms an enzyme-substrate complex $ZY$
to inhibit the functionality of $Z$ ($ZY+Y\rightleftharpoons ZYY$).
The interaction $Y\dashv Z\dashv Y$ functions as the PFL in this
network. In many cases, it is assumed that the concentration of enzyme
$Z$ equilibrates ($\dot{z}=0$), yielding the following two-dimensional
differential equation in terms of $\boldsymbol{v}=(x,y)^{\top}$,
where $x$ and $y$ denote the concentrations of mRNA $X$ and protein
$Y$, respectively: 
\begin{equation}
\boldsymbol{f}(\boldsymbol{v})=\left(\begin{array}{c}
{\displaystyle V_{sx}\frac{K_{d}^{p}}{K_{d}^{p}+y^{p}}-k_{dx}x}\\
{\displaystyle k_{sy}x-k_{dy}y-V_{dy}\frac{y}{K_{m}+y+y^{2}/K_{1}}}
\end{array}\right).\label{eq:F_relax_def}
\end{equation}
As in the smooth oscillator, we scaled the parameters so that the
period and the amplitude $y$ both became $1$ (Figure~\ref{fig:oscillation}B).
The parameters, $V_{sx}=0.815$, $K_{m}=0.0268$, $k_{dx}=3.04$,
$k_{sy}=60.8$, $k_{dy}=3.04$, $V_{dy}=16.3$, $K_{1}=0.134$, $K_{d}=0.268$,
and $p=4.0$, are essentially equivalent to those presented in the
Supplement of Nov\'ak and Tyson~\cite{Novak:2008:BiochemOsc}, except
for the scale transform ($V_{sx}=0.05$, $K_{m}=0.1$, $k_{dx}=0.05$,
$k_{sy}=1.0$, $k_{dy}=0.05$, $V_{dy}=1.0$, $K_{1}=0.5$, $K_{d}=1.0$,
and $p=4.0$ in the original model).

We analyzed two coupled relaxation oscillators, as shown in Figure~\ref{fig:motifs}F.
The coupling term $\boldsymbol{C}_{i}$ was written as a linear relation:
\begin{equation}
\boldsymbol{C}_{i}(\boldsymbol{v}_{i},\boldsymbol{v}_{j})=\left(\begin{array}{c}
\epsilon(x_{j}-x_{i})\\
0
\end{array}\right)\hspace{1em}((i,j)=(1,2)\,\,\mathrm{or}\,\,(2,1)),\label{eq:relax_coupling_term}
\end{equation}
where $\epsilon$ denotes the coupling strength.

\subsection{Analytical approaches}

For an analytical analysis, we used phase reduction and Floquet multipliers.
Even a system of only two coupled oscillators exhibits very complicated
behaviors, such as $n:m$ synchronization ($n,m$; positive integers)
and chaos, depending on the model parameters (see section~\ref{sec:Results}).
Because detailed bifurcation analysis of the coupled oscillator is
outside the scope of this paper, we assumed $1:1$ synchronization
between two oscillatory components (tight coupling). Under this assumption,
the coupled oscillator can be viewed as one oscillator, whose dynamics
is represented by 
\begin{equation}
\frac{d\boldsymbol{u}}{dt}=\boldsymbol{g}(\boldsymbol{u}),\label{eq:tilde_model}
\end{equation}
with 
\begin{equation}
\boldsymbol{u}=\left(\begin{array}{c}
\boldsymbol{v}_{1}\\
\boldsymbol{v}_{2}
\end{array}\right),\hspace{1em}\boldsymbol{g}(\boldsymbol{u})=\left(\begin{array}{c}
{\displaystyle \frac{\boldsymbol{f}(\boldsymbol{v}_{1})}{\tau_{1}}+\boldsymbol{C}_{1}(\boldsymbol{v}_{1},\boldsymbol{v}_{2})}\\
{\displaystyle \frac{\boldsymbol{f}(\boldsymbol{v}_{2})}{\tau_{2}}+\boldsymbol{C}_{2}(\boldsymbol{v}_{2},\boldsymbol{v}_{1})}
\end{array}\right).\label{eq:one_osc_map}
\end{equation}
Incorporating an input signal in equation~\ref{eq:tilde_model},
we have 
\begin{equation}
\frac{d\boldsymbol{u}}{dt}=\boldsymbol{g}(\boldsymbol{u})+\chi\boldsymbol{I}(\omega t),\label{eq:tilde_mode_input}
\end{equation}
where $\boldsymbol{I}(\omega t)$ denotes an input signal with angular
frequency $\omega$ and $\chi$ is a scalar representing the signal
strength. We will use equation~\ref{eq:tilde_mode_input} in the
calculations for entrainability and Floquet multipliers.

\subsubsection{Phase reduction}

\begin{figure}
\includegraphics[width=16.5cm]{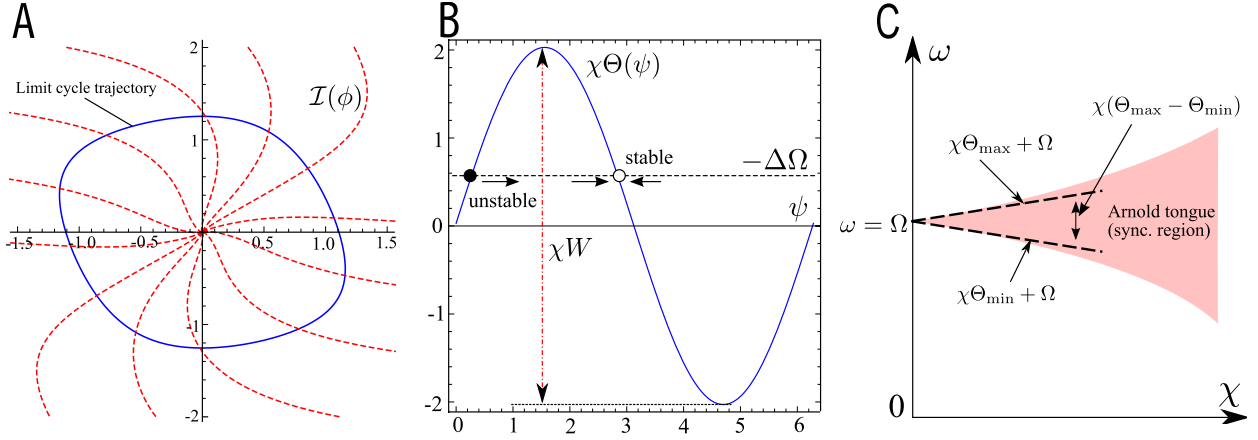}

\caption{(A) Trajectory of a hypothetical limit-cycle oscillator (solid line)
and its isochron $\mathcal{I}(\phi)$ (dashed lines), where the isochron
is drawn at intervals of $\pi/6$. (B) Illustration of $\dot{\psi}=\Delta\Omega+\chi\Theta(\psi)=0$,
where the solid and dashed line denote $\chi\Theta(\psi)$ and $-\Delta\Omega$,
respectively, for typical cases. A stable solution (empty circle)
exists inside $\chi\Theta_{\max}$ and $\chi\Theta_{\min}$, whose
length is denoted by $\chi W$. (C) Illustration of the relation between
the Arnold tongue (colored region), inside which the oscillator can
synchronize to an external signal (vertical and horizontal axes represent
the signal angular frequency $\omega$ and the signal strength $\chi$, respectively).
Equation~\ref{eq:sync_condition} (dashed line) is a linear approximation
of this region for sufficiently small $\chi$. The width of the region
can be approximated by $\chi(\Theta_{\max}-\Theta_{\min})$ around
$\chi=0$. \label{fig:isochron}}
\end{figure}

A phase-reduction approach uses a phase variable to reduce a multidimensional
system into a one-dimensional differential equation~\cite{Kuramoto:2003:OscBook,Izhikevich:2007:NeuroBook}.
In the absence of external perturbations (equation~\ref{eq:tilde_model}),
a limit-cycle oscillation can be described in terms of a phase variable
$\phi\in[0,2\pi]$: 
\begin{equation}
\frac{d\phi}{dt}=\Omega,\label{eq:phase_ODE}
\end{equation}
where $\Omega$ is the natural angular frequency of the limit cycle
($\Omega=2\pi/T$, where $T$ is the period). Although the phase $\phi$
in equation~\ref{eq:phase_ODE} is only defined on the stable limit-cycle
trajectory, its definition can be expanded to the entire $\boldsymbol{u}$
space, where the equiphase surface is called the isochron $\mathcal{I}(\phi)$.
An example of an isochron $\mathcal{I}(\phi)$ defined on a hypothetical
limit-cycle oscillator, shown by dashed lines, in Figure~\ref{fig:isochron}A
where the isochron is drawn at intervals of $\pi/6$.

In the presence of external signals (equation~\ref{eq:tilde_mode_input}),
the time evolution of $\phi$ is described by 
\begin{equation}
\frac{d\phi}{dt}=\Omega+\chi\boldsymbol{U}(\phi)\cdot\boldsymbol{I}(\omega t),\label{eq:phase_reduction_def}
\end{equation}
where a dot ($\cdot$) represents an inner product and $\boldsymbol{U}(\phi)$
is the phase response curve (PRC) defined by 
\begin{equation}
\boldsymbol{U}(\phi^{\prime})=\left.\nabla\phi(\boldsymbol{u})\right|_{\boldsymbol{u}=\boldsymbol{u}_{0}(\phi^{\prime})}.\label{eq:PRC_def}
\end{equation}
Here $\boldsymbol{u}_{0}(\phi^{\prime})$ represents a point on limit-cycle
oscillation at the phase $\phi^{\prime}$. It is assumed that the
perturbed trajectory is in the vicinity of the unperturbed limit-cycle
trajectory (i.e., $\chi$ is sufficiently small). The introduction
of a slow variable $\psi=\phi-\omega t$ in equation~\ref{eq:phase_reduction_def}
yields 
\begin{equation}
\frac{d\psi}{dt}=\Delta\Omega+\chi\boldsymbol{U}(\psi+\omega t)\cdot\boldsymbol{I}(\omega t),\label{eq:psi_ODE}
\end{equation}
where $\Delta\Omega=\Omega-\omega$. Because $\psi$ is a slow variable
($\dot{\psi}$ is very small), when $\chi$ is sufficiently small,
the inner product term can be approximated by its average (separating
the timescales of $\psi$ and $\boldsymbol{U}(\psi+\omega t)\cdot\boldsymbol{I}(\omega t)$):
\begin{eqnarray}
\frac{d\psi}{dt} & \simeq & \Delta\Omega+\chi\Theta(\psi),\label{eq:entrain_eq}
\end{eqnarray}
with 
\begin{equation}
\Theta(\psi)=\frac{1}{2\pi}\int_{0}^{2\pi}d\theta\,\boldsymbol{U}(\psi+\theta)\cdot\boldsymbol{I}(\theta),\label{eq:Theta_def}
\end{equation}
(the integration of equation~\ref{eq:Theta_def} is evaluated numerically
in practical calculations). If stable solutions exist, this oscillator
synchronizes to external signals. $\Theta(\psi)$ is a $2\pi$-periodic
function, and its shape is typically given by sinusoidal-like functions,
as shown in Figure~\ref{fig:isochron}B. The fixed points are the
intersection points of $\Theta(\psi)$ and $-\Delta\Omega$, and only
the point with $\frac{d\Theta}{d\psi}<0$ is a stable solution (an
empty circle in Figure~\ref{fig:isochron}B). Therefore, the oscillator
is synchronized to external signals if the following relation holds:
\begin{equation}
\chi\Theta_{\min}+\Omega<\omega<\chi\Theta_{\max}+\Omega,\label{eq:sync_condition}
\end{equation}
with 
\[
\Theta_{\min}=\min_{\psi}\Theta(\psi),\hspace{1em}\Theta_{\max}=\max_{\psi}\Theta(\psi).
\]
The extent of synchronization in terms of $\chi$ (signal strength)
and $\omega$ (signal angular frequency) is often described by a domain
known as the \emph{Arnold tongue} or the \emph{synchronization region}
\cite{Pikovsky:2001:SyncBook}, inside of which the oscillator can
synchronize to an external periodic signal (the colored domain in
Figure~\ref{fig:isochron}C). Although the border of the Arnold tongue
is generally nonlinear as a function of $\chi$, the upper and lower
borders can be approximated linearly by equation~\ref{eq:sync_condition}
when $\chi$ is sufficiently small (these are shown with the dashed
line in Figure~\ref{fig:isochron}C). Thus the range in which $\omega$
can be synchronized with an external signal can be quantified by $\chi(\Theta_{\max}-\Theta_{\min})$,
which represents the width of the Arnold tongue for sufficiently small
$\chi$. We define the entrainability $W$ as a $\chi$-independent
variable (see Figure~\ref{fig:isochron}B): 
\begin{equation}
W=\Theta_{\max}-\Theta_{\min}.\label{eq:entrainability_def}
\end{equation}
Writing equation~\ref{eq:sync_condition} in terms of the period
of the external signal, we have 
\begin{equation}
\frac{2\pi}{\chi\Theta_{\max}+\Omega}<T_{s}<\frac{2\pi}{\chi\Theta_{\min}+\Omega},\label{eq:sync_condition2}
\end{equation}
where $T_{s}$ is the period of the external signal ($T_{s}=2\pi/\omega$)
(note that equations~\ref{eq:sync_condition} and \ref{eq:sync_condition2}
only hold for sufficiently small $\chi$). The PRC {[}equation~\ref{eq:PRC_def}{]}
can be calculated by the following differential equation \cite{Izhikevich:2007:NeuroBook,Shwemmer:2012:WeaklyCoupleOsc}:
\begin{equation}
\frac{d\boldsymbol{U}(t)}{dt}=-\left\{ D\boldsymbol{g}(\boldsymbol{u}_{0}(t))\right\} ^{\top}\boldsymbol{U}(t),\label{eq:Malkin}
\end{equation}
with the constraint 
\begin{equation}
\boldsymbol{U}(t)\cdot\boldsymbol{g}(\boldsymbol{u}_{0}(t))=\Omega,\label{eq:PRC_constraint}
\end{equation}
where $\boldsymbol{u}_{0}(t)$ is a stable periodic solution of equation~\ref{eq:tilde_model}
($\boldsymbol{u}_{0}(t+T)=\boldsymbol{u}_{0}(t)$, where $T$ is the
period), $D\boldsymbol{g}(\boldsymbol{u}_{0}(t))$ is a Jacobian matrix
of $\boldsymbol{g}(\boldsymbol{u})$ around $\boldsymbol{u}_{0}(t)$,
and we abbreviated $\boldsymbol{U}(t)=\boldsymbol{U}(\phi(t))$ ($\phi(t)$
is  the phase as a function of $t$ according to equation~\ref{eq:phase_ODE}).
Because equation~\ref{eq:Malkin} is unstable, we numerically solved
equation~\ref{eq:Malkin} backward in time.

\subsubsection{Floquet multiplier}

In the presence of external signals, the solution $\boldsymbol{u}(t)$
deviates from a stable orbit $\boldsymbol{u}_{0}$, where the deviation
$\boldsymbol{\eta}=\boldsymbol{u}-\boldsymbol{u}_{0}$ obeys the differential
equation 
\begin{equation}
\frac{d\boldsymbol{\eta}(t)}{dt}=D\boldsymbol{g}(\boldsymbol{u}_{0}(t))\boldsymbol{\eta}(t).\label{eq:sol}
\end{equation}
Here we assume that the deviation is sufficiently small. Because $D\boldsymbol{g}(\boldsymbol{u}_{0})$
is a periodic function with period $T$, equation~\ref{eq:sol} has
the Floquet solutions, generally represented by 
\begin{equation}
\boldsymbol{\eta}(t)=\sum_{i=1}^{N}c_{i}\exp(\mu_{i}t)\boldsymbol{p}_{i}(t),\label{eq:Floquet_solution}
\end{equation}
where $\boldsymbol{p}_{i}(t)$ are functions with period $T$, $c_{i}$
is a coefficient determined by the initial conditions, and $\mu_{i}$
are the Floquet exponents. The Floquet multipliers $\rho_{i}$ are
related to the exponents via 
\begin{equation}
\rho_{i}=\exp(\mu_{i}T).\label{eq:Floquet_multiplier}
\end{equation}
Note that the Floquet exponents are a special case of the Lyapunov
exponents for periodic systems (e.g., limit cycles) \cite{Klausmeier:2008:Floquet}.
The Floquet multipliers can be calculated numerically from a matrix
differential equation in terms of the matrix $Q(t)$, after Klausmeier~\cite{Klausmeier:2008:Floquet}:
\begin{equation}
\frac{dQ(t)}{dt}=D\boldsymbol{g}(\boldsymbol{u}_{0}(t))Q(t).\label{eq:func_ode}
\end{equation}
Equation~\ref{eq:func_ode} is integrated numerically from $t=0$
to $t=T$ with an initial condition that $Q(t=0)$ is the identity
matrix. The Floquet multipliers are obtained as the eigenvalues of
$Q(t=T)$, and hence $N$-dimensional limit-cycle models yield $N$
Floquet multipliers.

When the Floquet multipliers $\rho_{i}$ cross the unit circle $|\rho|=1$
on the complex plane, the limit cycle bifurcates. Bifurcations of
codimension 1 can occur in three patterns \cite{Guckenheimer:1983:DynsysBook,Crawford:1991:IntroBifTheory,Wiesenfeld:1985:PDBifAmp,Wiesenfeld:1986:SignalAmpBif}: 
\begin{itemize}
\item Saddle-node bifurcation: A single real Floquet multiplier crosses
$\rho=1$ (this includes transcritical or pitchfork bifurcations as
special cases); 
\item Hopf bifurcation: Complex-conjugate Floquet multipliers cross $\left|\rho\right|=1$
($\rho\ne\pm1$); 
\item Period-doubling bifurcation: A single real Floquet multiplier crosses
$\rho=-1$. 
\end{itemize}
In autonomous systems, one of the Floquet multipliers is $\rho_{i}=1$,
which corresponds to a pure phase shift. Wiesenfeld and McNamara showed
that Floquet multipliers whose absolute values are close to 1 are
responsible for signal amplification~\cite{Wiesenfeld:1985:PDBifAmp,Wiesenfeld:1986:SignalAmpBif}.
We calculated the Floquet multipliers up to the second-largest absolute
value (except for the inherent multiplier $\rho=1$), which are denoted
as $\rho_{1}$ (the largest) and $\rho_{2}$ (the second largest)
in our models.

\section{Results\label{sec:Results}}

We performed a bifurcation analysis and examined the relationship
between entrainability and the Floquet multipliers. We set $\tau_{1}=1$
and $\tau_{2}=R$ in equations~\ref{eq:coupled_1} and \ref{eq:coupled_2},
where $R$ represents the ratio of the periods of the two oscillators.
When $R\ne1$, the two oscillatory components have different natural
periods (without coupling) and hence the periods are mismatched. As
the controllable parameters for the model calculations, we adopted
$R$ and $\epsilon$, where $\epsilon$ denotes the coupling strength
defined in equations~\ref{eq:smooth_coupling_term} and \ref{eq:relax_coupling_term}.

\subsection{Bifurcation diagram}

Bifurcations of the limit cycle are classified into bifurcations in
equilibria and those on the limit cycle. Here, bifurcation on the
limit cycle was studied by numerically solving differential equations~\ref{eq:F_smooth_def}
and \ref{eq:F_relax_def}.

\begin{figure}
\includegraphics[width=15cm]{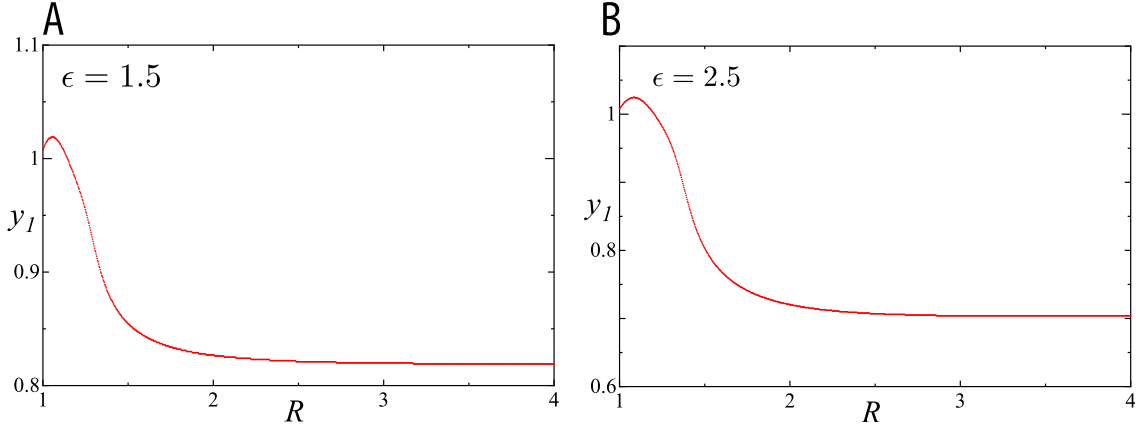}

\caption{Bifurcation diagram (local maxima of $y_{1}$) of the smooth oscillator
as a function of $R$, where (A) $\epsilon=1.5$, and (B) $\epsilon=2.5$.
\label{fig:bif_smooth_R}}
\end{figure}

Specifically, we described the bifurcation diagrams as a function
of $R$ (period ratio) and $\epsilon$ (coupling strength) and calculated
the local maxima of $y_{1}$. If the oscillator showed regular limit-cycle
oscillations, the maxima of $y_{1}$ was represented by a single point.
If the limit cycle underwent bifurcation, the maximal points were
described by more than two points. We iterated by changing $R$ from
$R_{\min}$ to $R_{\max}$ ($R_{\min}<R_{\max}$) and then adopted
the last values of the preceding $R$ values as the initial values
for the next $R$ value. The incremental step for $R$ was set at
$\Delta R=(R_{\max}-R_{\min})/1000$, where $R_{\min}$ and $R_{\max}$
are indicated as the lower and upper limits on Figures~\ref{fig:bif_smooth_R}
and \ref{fig:bif_R_relax}. At each parameter value, we discarded
as transient phases those during $t\in[0,200]$, and we recorded the
local maxima of the trajectories in during the intervals $t\in[200,300]$.
We carried out these same procedures on the coupling parameter $\epsilon$
($\epsilon_{\min}$, $\epsilon_{\max}$, with the increment $\Delta\epsilon=(\epsilon_{\max}-\epsilon_{\min})/1000$).

\begin{figure}
\includegraphics[width=15cm]{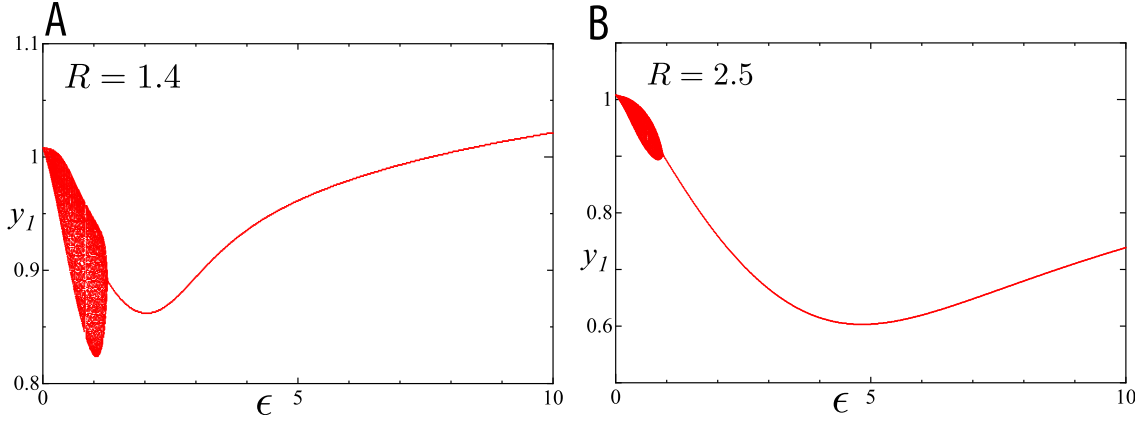}

\caption{Bifurcation diagram (local maxima of $y_{1}$) of the smooth oscillator
as a function of $\epsilon$, where (A) $R=1.4$, and (B) $R=2.5$.
\label{fig:bif_smooth_epsilon}}
\end{figure}

\begin{figure}
\includegraphics[width=15cm]{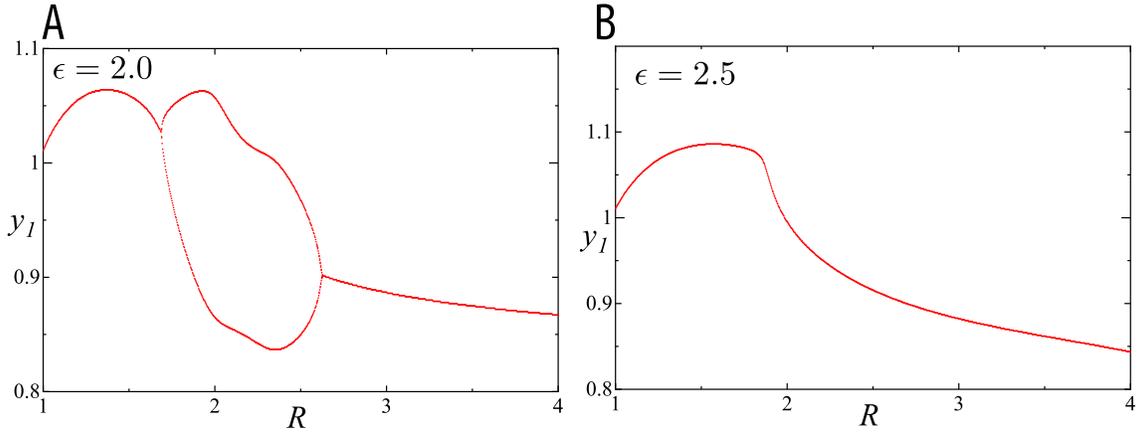}

\caption{Bifurcation diagram (local maxima of $y_{1}$) of the relaxation oscillator
as a function of $R$, where (A) $\epsilon=2.0$, and (B) $\epsilon=2.5$.
\label{fig:bif_R_relax}}
\end{figure}

In Figure~\ref{fig:bif_smooth_R}, the bifurcation diagram of $y_{1}$
as a function of $R$ in the smooth oscillator is shown for (A) $\epsilon=1.5$
and (B) $\epsilon=2.5$. Although the maxima of $y_{1}$ slightly
depend on $R$, there is no bifurcation. In Figure~\ref{fig:bif_smooth_epsilon},
the bifurcation diagram as a function of $\epsilon$ is shown for
(A) $R=1.4$ and (B) $R=2.5$. In both cases, bifurcations occurred
at $\epsilon\simeq1.3$ for $R=1.4$ and $\epsilon\simeq0.9$ for
$R=2.5$, and chaotic behavior was observed when the coupling strength
was small.

\begin{figure}
\includegraphics[width=15cm]{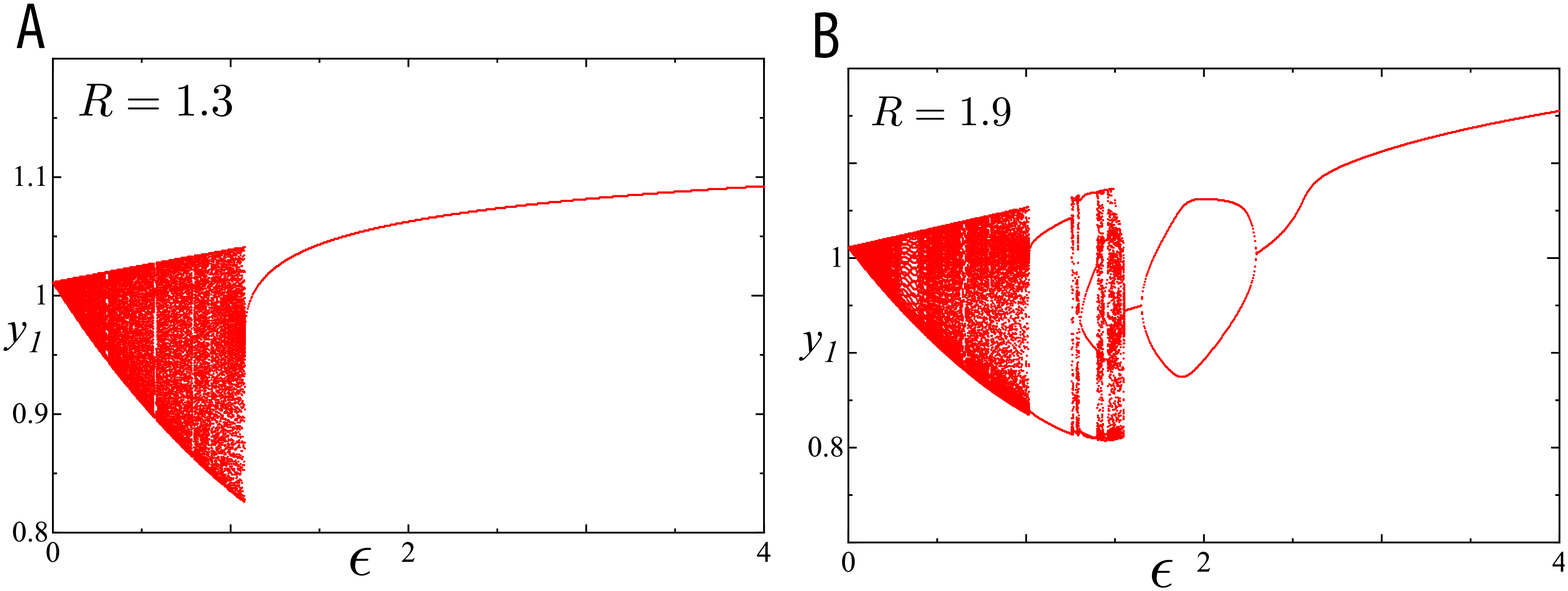}

\caption{Bifurcation diagram (local maxima of $y_{1}$) of the relaxation oscillator
as a function of $\epsilon$, where (A) $R=1.3$, and (B) $R=1.9$.
\label{fig:bif_epsilon_relax}}
\end{figure}

Figure~\ref{fig:bif_R_relax} shows the bifurcation diagram as a
function of $R$ for the relaxation oscillator for (A) $\epsilon=2.0$
and (B) $\epsilon=2.5$. In the former case, a period-doubling bifurcation
occurred in the region $1.7<R<2.7$. In the latter, a cusp appeared
around $R=1.9$, in the vicinity of a saddle-node bifurcation (see
the section on entrainability and Floquet multipliers). In Figure~\ref{fig:bif_epsilon_relax},
the bifurcation diagram as a function of $\epsilon$ is shown for
(A) $R=1.3$ and (B) $R=1.9$. A complex behavior (chaos) appeared
in a small region of $\epsilon$. Especially in Figure~\ref{fig:bif_epsilon_relax}B,
we found a branch of a maxima around $1.7<\epsilon<2.3$, indicating
the occurrence of the period-doubling bifurcation.

In comparison, bifurcations are less likely in a smooth oscillator
than in a relaxation oscillator; bifurcations are only observed in
the case of very weak coupling in smooth oscillators (Figure~\ref{fig:bif_smooth_epsilon}).
In contrast, the relaxation oscillator exhibited period-doubling and
chaotic oscillations in a wider range of coupling strengths (Figures
~\ref{fig:bif_R_relax} and \ref{fig:bif_epsilon_relax}).

\begin{figure}
\includegraphics[width=15cm]{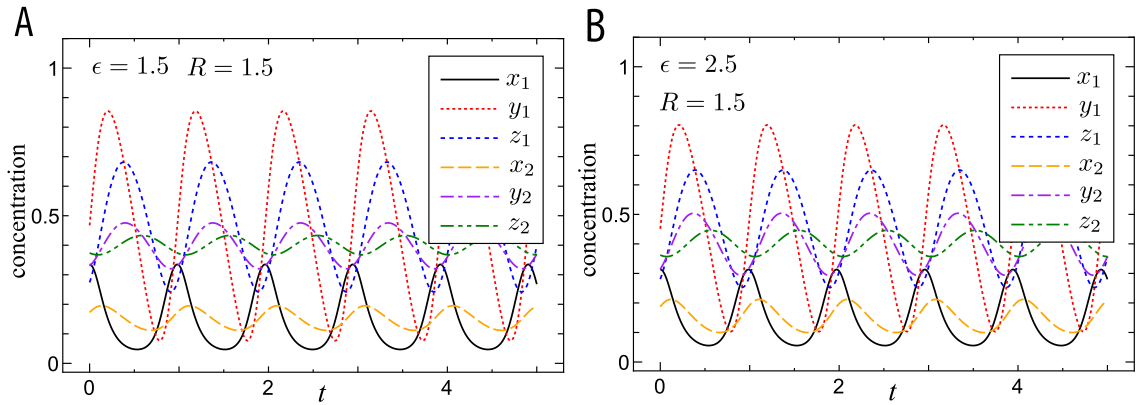}

\caption{Time course of the coupled smooth oscillator with (A) $\epsilon=1.5$
and $R=1.5$, and (B) $\epsilon=2.5$ and $R=1.5$. \label{fig:timecource_couple_smooth}}
\end{figure}

\begin{figure}
\includegraphics[width=15cm]{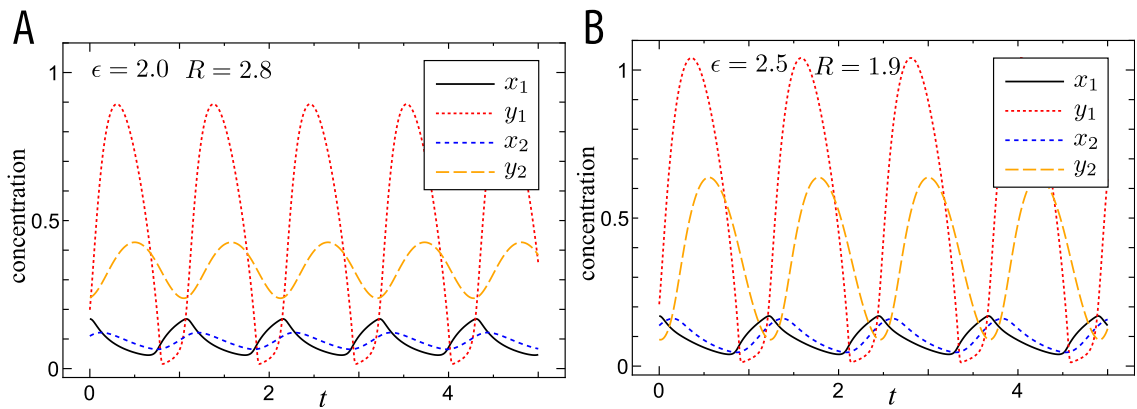}

\caption{Time course of the coupled relaxation oscillator, with (A) $\epsilon=2.0$
and $R=2.8$, and (B) $\epsilon=2.5$ and $R=1.9$. \label{fig:timecourse_couple_relax}}
\end{figure}

Next, we examine the time course of the coupled oscillators with mismatched
periods (Figures~\ref{fig:timecource_couple_smooth} and \ref{fig:timecourse_couple_relax},
respectively, for the smooth and relaxation oscillators). Without
mismatched periods, the temporal trajectory of the oscillators would
be identical to that of the case of a single oscillator (Figure~\ref{fig:oscillation}).
Figure~\ref{fig:timecource_couple_smooth} shows the case for coupled
smooth oscillators with (A) $\epsilon=1.5$, $R=1.5$, and (B) $\epsilon=2.5$,
$R=1.5$; Figure~\ref{fig:timecourse_couple_relax} shows the case
for coupled relaxation oscillators with (A) $\epsilon=2.0$, $R=2.8$,
and (B) $\epsilon=2.5$, $R=1.9$. In all the cases shown in Figures~\ref{fig:timecource_couple_smooth}
and \ref{fig:timecourse_couple_relax}, we see the $1:1$ synchronization
between the two oscillators. Without period mismatch, the amplitude
of oscillation $y$ is $1$ (see Methods). We see from Figures~\ref{fig:timecource_couple_smooth}
and \ref{fig:timecourse_couple_relax} that the amplitudes of $y_{1}$
and $y_{2}$ depend on the period ratio $R$, where the amplitude
of $y_{1}$ is smaller than 1 except in Figure~\ref{fig:timecourse_couple_relax}B.
From Figures~\ref{fig:bif_smooth_R} and \ref{fig:bif_R_relax},
which describe maxima of $y_{1}$ as a function $R$, we see that
$y_{1}$ is larger than 1 in both oscillator models when the period
mismatch is smaller, and the maxima of $y_{1}$ start to decrease
with increasing $R$. The decrease of the amplitude is larger for
the smooth oscillator than for the relaxation oscillator.

\subsection{Entrainability and Floquet multiplier}

In using phase reduction and Floquet multipliers to calculate how
the entrainability $W$ (equation~\ref{eq:entrainability_def}) depends
on $R$ (equation~\ref{eq:R_def}) and $\epsilon$ (equations~\ref{eq:smooth_coupling_term}
and \ref{eq:relax_coupling_term}), we employed the following sinusoidal
input signal: 
\begin{equation}
\boldsymbol{I}(\omega t)=\left(\begin{array}{c}
\sin(\omega t)\\
0\\
0\\
\vdots
\end{array}\right),\label{eq:input_signal}
\end{equation}
where the input signal only affects the concentration of mRNA in the
first oscillator ($x_{1}$ in equation~\ref{eq:tilde_model}). We
assumed $1:1$ synchronization between the two oscillators (the regions
where the maximum of $y_{1}$ is represented by a single point in
Figures~\ref{fig:bif_smooth_R}--\ref{fig:bif_epsilon_relax}) so
that the coupled oscillator could be viewed as a single case. Although
we calculated the entrainability and the Floquet multipliers inside
the $1:1$ synchronized regions, the entrainability very close to
the bifurcation points is not shown for some parameters, because equation~\ref{eq:Malkin}
did not exhibit stable periodic solutions even after remaining there
for a long time ($\epsilon\simeq1.3$ in Figure~\ref{fig:smooth_as_eps}A,
$R\simeq2.7$ in Figure~\ref{fig:relax_as_R}A and $\epsilon\simeq2.3$
in Figure~\ref{fig:relax_as_eps}D). In most cases we calculated
the Floquet multipliers up to the second most dominant ones. When
the multipliers were given by complex-conjugate values, we checked
the three multipliers $\rho_{1}$, $\bar{\rho}_{1}$, and $\rho_{2}$,
where $\bar{\rho}$ is the complex conjugate of $\rho$. When two
real multipliers ($\rho_{1}$ and $\rho_{1}^{\prime}$) eventually
emerged due to annihilation of the complex-conjugate multipliers ($\rho_{1}$
and $\bar{\rho}_{1}$) by changing parameters, we checked the three
multipliers $\rho_{1}$, $\rho_{1}^{\prime}$, and $\rho_{2}$.

By calculating the entrainability $W$ and the Floquet multipliers
$\rho_{i}$, we obtained stable limit-cycle trajectories in a way
similar to what we did in the bifurcation analysis. We iterated over
$R$ and adopted the values of the preceding $R$ values as the initial
values for the next $R$ values. For each parameter value, we calculated
the stable limit-cycle trajectories for one period, and these trajectories
were used for the calculation of the PRC functions (Equation~\ref{eq:PRC_def}).
These procedures were also carried out for the coupling parameter
$\epsilon$. The Floquet multipliers also require stable limit-cycle
trajectories, and we calculated them in the same way as described
above.

\subsubsection{Smooth oscillator}

\begin{figure}
\includegraphics[width=15cm]{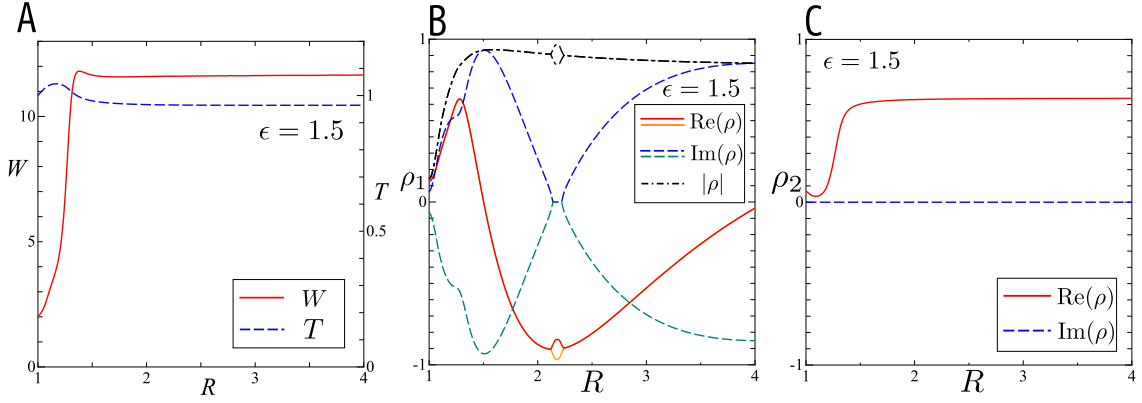}

\caption{Entrainability $W$, period $T$, and Floquet multipliers $\rho_{i}$,
as a function of $R$ in the smooth oscillator. (A--C) $R$ dependence
of (A) $W$ and $T$, (B) $\rho_{1}$, and (C) $\rho_{2}$, for $\epsilon=1.5$.
In (A), the solid and dashed lines denote the entrainability and the
period, respectively. In B and C, the solid, dashed, and dot-dashed
lines denote the real part, the imaginary part, and absolute value
of the Floquet multipliers, respectively. \label{fig:smooth_as_R}}
\end{figure}

\begin{figure}
\includegraphics[width=15cm]{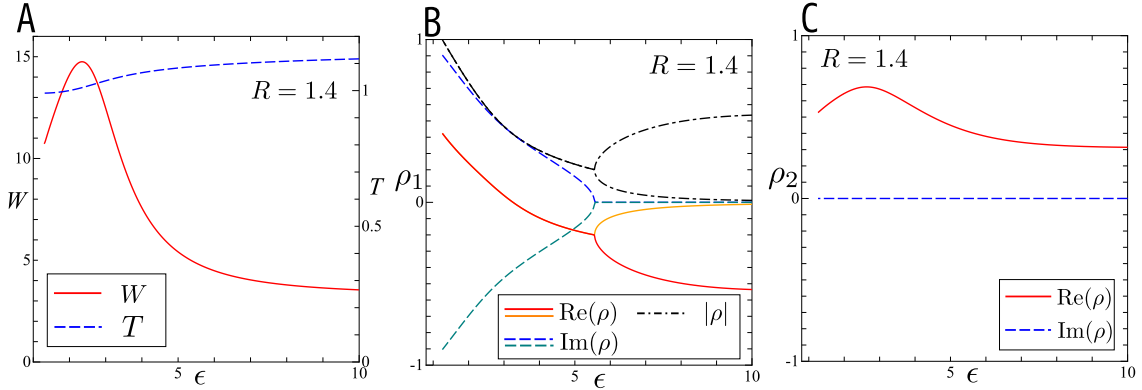}

\caption{Entrainability $W$, period $T$, and Floquet multipliers $\rho_{i}$,
as a function of $\epsilon$ in the smooth oscillator. (A--C) $\epsilon$
dependence of (A) $W$ and $T$, (B) $\rho_{1}$, and (C) $\rho_{2}$,
for $R=1.4$. The meanings of the lines are the same as in Figure~\ref{fig:smooth_as_R}.
\label{fig:smooth_as_eps}}
\end{figure}

We first investigated the smooth oscillator as a function of $R$
(Figure~\ref{fig:smooth_as_R}). Specifically, Figure~\ref{fig:smooth_as_R}A
is a dual-axis plot of the entrainability $W$ (solid line; left axis)
and the period $T$ (dashed line; right axis) for $\epsilon=1.5$.
We can uniquely define the period $T$ of the coupled oscillator because
we are considering a $1:1$ synchronization between the two oscillators.
In Figure~\ref{fig:smooth_as_R}A, the period $T$ is not strongly
affected by period mismatch and is around $1$ in the whole $R$ region,
although the entrainability $W$ depends on $R$. The entrainability
is $W\simeq2$ when the two oscillators have the same period ($R=1$),
whereas $W\simeq12$ for $R\geq1.4$. The first and the second dominant
Floquet multipliers for $\epsilon=1.5$ are shown in Figure~\ref{fig:smooth_as_R}B
and C, where solid, dashed, and dot-dashed lines respectively denote
$\mathrm{Re}(\rho_{i})$, $\mathrm{Im}(\rho_{i})$, and $|\rho_{i}|$
($\rho_{i}$: Floquet multiplier defined by equation~\ref{eq:Floquet_multiplier}).
In Figure~\ref{fig:smooth_as_R}B, symmetric imaginary curves represent
complex-conjugate multipliers, and a branch of the real part around
$R\simeq2.2$ is due to annihilation of the complex-conjugate multipliers
(two real multipliers emerge in place of the complex-conjugate multipliers).
We see that the $\rho_{1}$ in Figure~\ref{fig:smooth_as_R}B are
imaginary values, and their absolute values are close to 1 for the
$R>1.4$ region. The coupled oscillator is close to the Hopf bifurcation
points. The second dominant Floquet multiplier $\rho_{2}$ is described
in Figure~\ref{fig:smooth_as_R}C, and we see that its magnitude
is comparable to that of $\rho_{1}$. The coupled oscillator is near
the saddle-node bifurcation points (the second Floquet multiplier
$\rho_{2}$ is a pure real value). We also carried out the same calculations
with $\epsilon=2.5$ and found enhanced entrainability $W\simeq16$
in the presence of a period mismatch ($R>1.5$), as in Figure~\ref{fig:smooth_as_R}C
(data not shown). According to the analysis with Floquet multipliers,
this enhancement can also be accounted for by the bifurcation.

Figure~\ref{fig:smooth_as_eps} shows the entrainability and Floquet
multipliers as a function of $\epsilon$ in the smooth oscillator
while keeping $R$ constant. Figure~\ref{fig:smooth_as_eps}A is
a dual-axis plot of the entrainability $W$ (left axis) and the period
$T$ (right axis) for $R=1.4$, where the period $T$ does not strongly
depend on the coupling strength $\epsilon$. In contrast, the entrainability
$W$ achieves a maximum, $W\simeq15$ at $\epsilon\simeq2.5$, which
is more than 7 times larger than the no-mismatch case ($W\simeq2$).
Figure~\ref{fig:smooth_as_eps}B shows the first dominant Floquet
multiplier $\rho_{1}$ as a function of $\epsilon$, where the notation
is the same as in Figure~\ref{fig:smooth_as_R}. $\rho_{1}$ is a
complex multiplier for $\epsilon<5.5$, and its absolute value (dot-dashed
line) approaches $1$ as $\epsilon$ decreases, showing that the oscillator
is in the vicinity of the Hopf bifurcation. We see consistent results
in Figure~\ref{fig:bif_smooth_epsilon} where the oscillator is quasiperiodic
(chaotic) in the $\epsilon<1.3$ region. At $\epsilon\simeq5.5$,
the complex-conjugate multipliers annihilate, and two real multipliers
emerge instead. Figure~\ref{fig:smooth_as_eps}C shows the second
dominant Floquet multiplier $\rho_{2}$ as a pure positive real number,
responsible for the saddle-node bifurcation points (this multiplier
is close to $\rho_{i}=1$). We see that $\rho_{2}$ achieves its maximal
value around $\epsilon\simeq2.5$, where the entrainability $W$ also
exhibits a maximum. Because $|\rho_{1}|$ in Figure~\ref{fig:smooth_as_eps}B
monotonically decreases as a function of $\epsilon$ for $\epsilon<5.5$,
$\rho_{1}$ does not explain the qualitative behavior of $W$, whereas
$\rho_{2}$ does. This result shows that being close to the saddle-node
bifurcation is important for achieving entrainment. We investigated
the entrainability with a different parameter for the period mismatch
$R=2.5$, and observed that $W\simeq26$ was maximized at an intermediate
coupling strength $\epsilon\simeq5$ (data not shown). The Floquet
multiplier analysis also indicated that enhancement in the $R=2.5$
case is caused by the oscillators being in the vicinity of the saddle-node
bifurcation on the limit cycle.

\subsubsection{Relaxation oscillator}

\begin{figure}
\includegraphics[width=15cm]{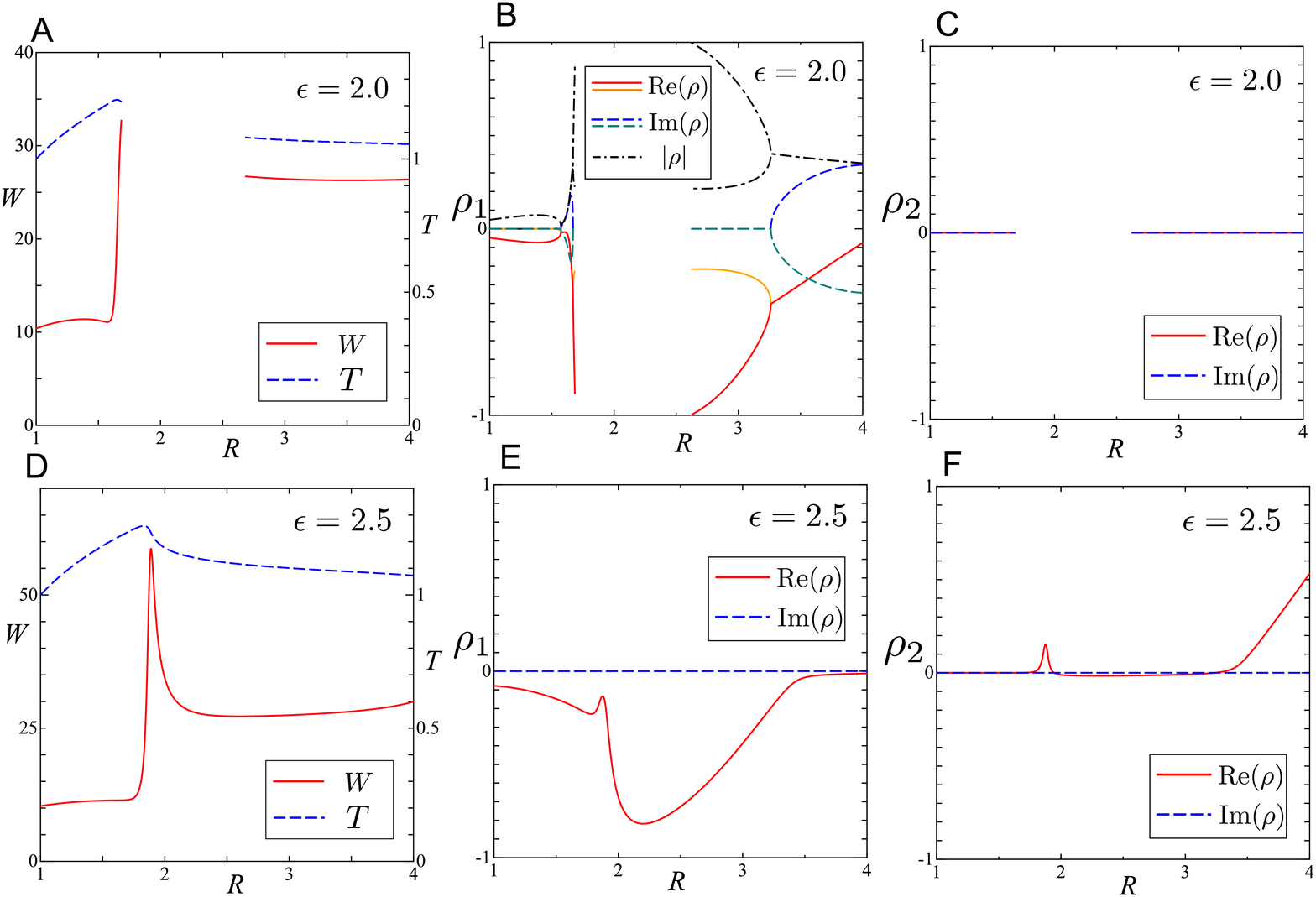}

\caption{Entrainability $W$, period $T$, and Floquet multipliers $\rho_{i}$,
as a function of $R$ for the relaxation oscillator. (A--C) $R$ dependence
of (A) $W$ and $T$, (B) $\rho_{1}$, and (C) $\rho_{2}$ for $\epsilon=2.0$.
(D--F) $R$ dependence of (D) $W$ and $T$, (E) $\rho_{1}$, and
(F) $\rho_{2}$ for $\epsilon=2.5$. The meanings of lines are the
same as in Figure~\ref{fig:smooth_as_R}. \label{fig:relax_as_R}}
\end{figure}

\begin{figure}
\includegraphics[width=15cm]{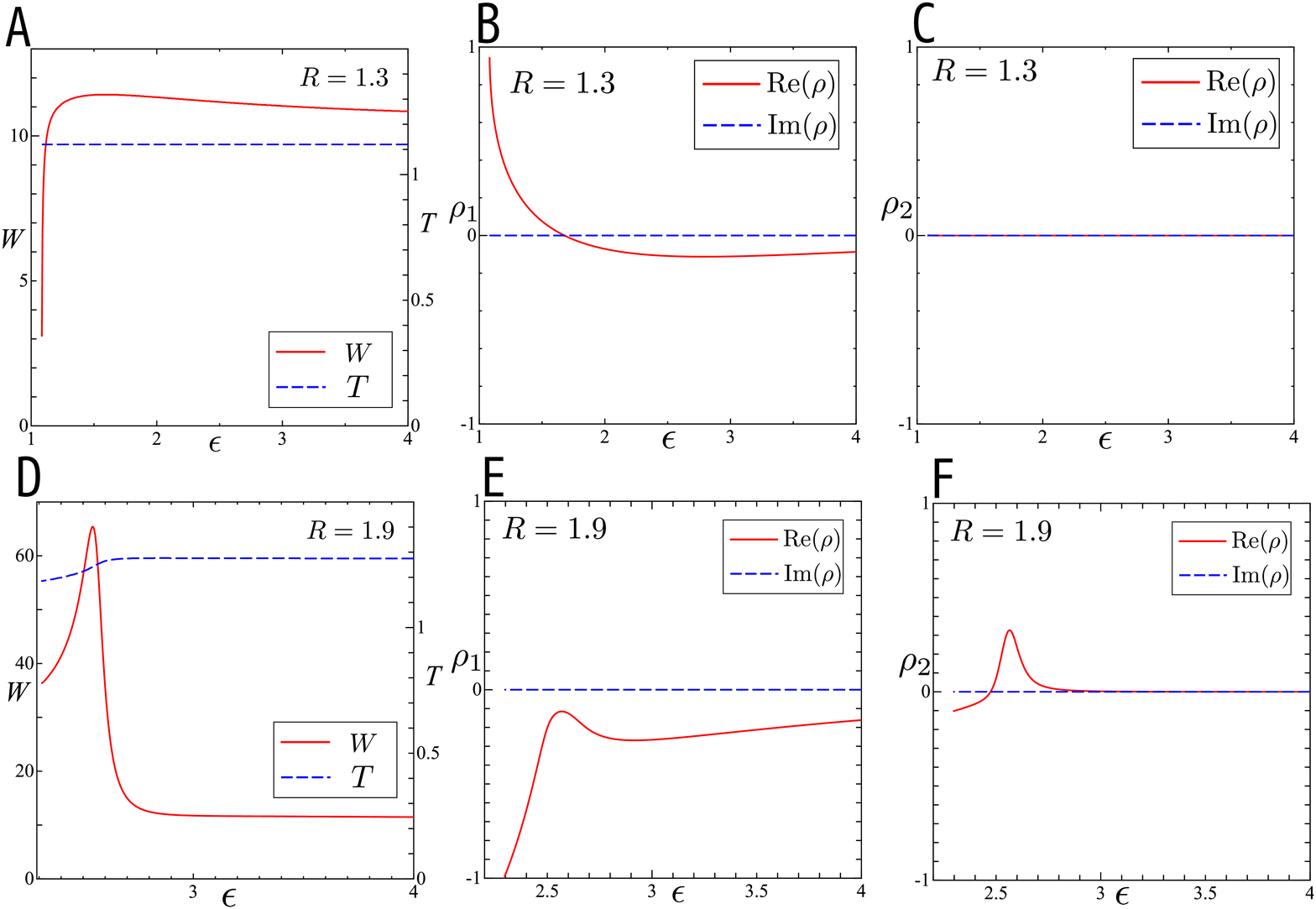}

\caption{Entrainability $W$, period $T$, and Floquet multipliers $\rho_{i}$,
as a function of $\epsilon$ for the relaxation oscillator. (A--C)
$\epsilon$ dependence of (A) $W$ and $T$, (B) $\rho_{1}$, and
(C) $\rho_{2}$, for $R=1.3$. (D--F) $\epsilon$ dependence of (D)
$W$ and $T$, (E) $\rho_{1}$, and (F) $\rho_{2}$, for $R=1.9$.
The meanings of the lines are the same as in Figure~\ref{fig:smooth_as_R}.
\label{fig:relax_as_eps}}
\end{figure}

We next apply the same analysis to the relaxation oscillator: we calculated
the entrainability $W$, the period $T$, and the Floquet multipliers
$\rho_{i}$, as functions of $R$ (Figure~\ref{fig:relax_as_R})
and $\epsilon$ (Figure~\ref{fig:relax_as_eps}). Figure~\ref{fig:relax_as_R}A
shows a dual-axis plot of the entrainability $W$ and the period $T$,
for $\epsilon=2.0$ (see also Figure~\ref{fig:smooth_as_R}). The
coupled oscillator is not $1:1$ synchronized in the region $1.7<R<2.7$,
and these unsynchronized regions were not plotted. Since the entrainability
$W$ grows rapidly in the vicinity of the bifurcation points ($R\simeq1.7$)
in Figure~\ref{fig:relax_as_R}A, we calculated the first $\rho_{1}$
and second $\rho_{2}$ dominant Floquet multipliers in Figure~\ref{fig:relax_as_R}B
and C, where $\rho_{1}$ is a complex-conjugate multiplier around
$R=1.7$ and is a pure real multiplier very near the bifurcation points.
We see that $\rho_{1}$ of $\epsilon=2.0$ exhibits a sudden decrease
towards $-1$ near $R\simeq1.7$ and $R\simeq2.7$, indicating that
the coupled oscillator is close to the period-doubling bifurcation
points. The bifurcation diagram of Figure~\ref{fig:bif_R_relax}
shows a branch, i.e., the period-doubling bifurcation. $\rho_{1}$
is a real multiplier for $R<3.3$, and its value approaches $-1$
as $R$ decreases toward $R=2.7$ in the vicinity of the period-doubling
bifurcation. At $R\simeq3.3$, complex-conjugate multipliers emerge
instead of the annihilation of two real multipliers. Figure~\ref{fig:relax_as_R}C
describes the second dominant Floquet multiplier $\rho_{2}$ of $\epsilon=2.0$,
and both the imaginary and real parts of $\rho_{2}$ vanish for a
range of $R$ values. Because the rest of the Floquet multipliers
for $\epsilon=2.0$ vanish (not shown here), all the dominant Floquet
multipliers of $\epsilon=2.0$ are related to the period-doubling
bifurcation points.

Figure~\ref{fig:relax_as_R}D shows the entrainability $W$ and the
period $T$ for $\epsilon=2.5$. We also see the rapid increase in
$W$ around $R\simeq1.9$, and the period $T$ increases up to $T=1.3$
for $1\le R<1.9$ and then decreases for $R>1.9$. Figures~\ref{fig:relax_as_R}E
and F show the first $\rho_{1}$ and the second $\rho_{2}$ dominant
Floquet multipliers for $\epsilon=2.5$, respectively. $\rho_{1}$
is a pure real multiplier with a minimum around $R\simeq2.2$, which
does not coincide with the position of the maximum of $W$. Although
the magnitude of $\rho_{2}$ is smaller than $\rho_{1}$, it gains
a peak at $R\simeq1.9$ identical to the peak position of $W$. Thus,
the system is better entrained in the vicinity of the saddle-node
bifurcation, and there exists an optimal coupling strength for the
entrainability.

We next show the entrainability $W$, the period $T$, and the Floquet
multipliers $\rho_{i}$ as a function of $\epsilon$ for the relaxation
oscillator. Figure~\ref{fig:relax_as_eps}A is a dual-axis plot of
the entrainability $W$ and the period $T$ for $R=1.3$. We see very
little dependence of the entrainability or the period on $\epsilon$.
Figures~\ref{fig:relax_as_eps}B and C show the first $\rho_{1}$
and the second $\rho_{2}$ dominant Floquet multipliers, respectively;
$\rho_{1}$ approaches $1$ with decreasing $\epsilon$, indicating
that the oscillator approaches the saddle-node bifurcation points,
whereas $\rho_{2}$ vanishes for a range of values of $\epsilon$.
Figure~\ref{fig:relax_as_eps}D shows the entrainability $W$ and
the period $T$ as a function of $\epsilon$ for $R=1.9$, where the
period $T$ does not strongly depend on the coupling strength $\epsilon$.
In contrast, the entrainability $W$ exhibits nonmonotonic behavior
and achieves the local maximum around $\epsilon=2.5$. The first dominant
Floquet multiplier $\rho_{1}$ in Figure~\ref{fig:relax_as_eps}E
shows that the oscillator approaches the period-doubling bifurcation
when approaching $\epsilon=2.3$ from above. Although the first dominant
Floquet multiplier shows the period-doubling bifurcation, its magnitude
does not qualitatively explain the existence of the entrainability
peak around $\epsilon=2.5$. In contrast, the second dominant Floquet
multiplier $\rho_{2}$ in Figure~\ref{fig:relax_as_eps}F shows a
positive-valued peak around $\epsilon=2.5$, and the position agrees
with that of the entrainability. In Figure~\ref{fig:relax_as_eps}D,
enhancement is only observed inside a narrow region where $\epsilon<2.8$.

\section{Discussion}

Mismatched periods enhance entrainability for both smooth and relaxation
oscillators, and the enhancement is related to the dominant Floquet
multipliers. The behavior as a function of the model parameters ($R$
and $\epsilon$) can be explained qualitatively by a pure real Floquet
multiplier. It has been previously shown that in the vicinity of the
bifurcation points, the oscillator is more sensitive to an external
signal \cite{Wiesenfeld:1985:PDBifAmp,Wiesenfeld:1986:SignalAmpBif}.
Entrainability is thus enhanced because the period-mismatch drives
the coupled oscillator to the vicinity of saddle-node bifurcation.
The pure real multiplier has a strong impact on the entrainment property
(Figures~\ref{fig:smooth_as_R}, \ref{fig:smooth_as_eps}, \ref{fig:relax_as_R}D--F
and \ref{fig:relax_as_eps}D--F), as do the Floquet multipliers related
to other bifurcation types (Figures~\ref{fig:relax_as_R}A--C). In
Figure~\ref{fig:relax_as_R}A, we see that even when neither the
first nor the second dominant Floquet multiplier is a positive real
multiplier, enhancement is induced by the period-doubling bifurcation
points. 

From our analysis above, we saw that bifurcation is less likely to
occur in a smooth oscillator than in a relaxation oscillator. Nevertheless,
the enhancement of entrainability (an acute peak for a relaxation
oscillator and an obtuse peak for a smooth oscillator) is seen in
a wider region for a smooth oscillator than for a relaxation oscillator
(Figures~\ref{fig:smooth_as_R}A and \ref{fig:relax_as_R}D for $R$
dependence, and Figures~\ref{fig:smooth_as_eps}A and \ref{fig:relax_as_eps}D
for $\epsilon$ dependence). Although a smooth oscillator may be better
entrained, Figures~\ref{fig:timecource_couple_smooth} and \ref{fig:timecourse_couple_relax}
show that the amplitude of the oscillation is more strongly affected
in the smooth oscillator. When increasing the period mismatch $R$,
the amplitude of $y_{1}$ increases at the outset and then starts
to decrease; the decrease starts sooner for the smooth oscillator
than it does for the relaxation oscillator (see Figure~\ref{fig:bif_smooth_R}
and \ref{fig:bif_R_relax}). Since limit-cycle oscillations with smaller
amplitude are more sensitive to external stimuli, the reduction of
the amplitude that is induced by the period mismatch is also a cause
of the improved entrainability. However, the main contributing factor
to the enhancement is the bifurcation on the limit cycle. This is
because the behavior of the entrainability $W$ agreed with that of
the Floquet multiplier, which does not depend on the scaling of the
values.

As mentioned in the introduction, many circadian oscillators consist
of multioscillatory networks, each of which has different components.
As an example, early works on \emph{Gonyaulax} reported periodic bioluminescence
($24$ h to $27$ h: average $25$ h) and aggregation ($15$ h to
$23$ h: average $20$ h) patterns~\cite{Roenneberg:1993:TwoOsc,Morse:1994:TwoCircadian}.
These two oscillators are synchronized in the presence of white light,
whereas they are decoupled in the presence of dim red light. The ratio
of the two periods is about $1.25$ (see Figure~2(c) in Roenneberg~\cite{Roenneberg:1993:TwoOsc}),
which coincides with our results of $R\simeq1.4$ for enhancement
in the smooth oscillator (see Figure~\ref{fig:smooth_as_R}A).

We have shown that the oscillators are better synchronized to an external
signal when they are close to bifurcation points (mainly at the saddle-node
but also partially at the period-doubling bifurcations). Regarding
the period-doubling bifurcation in realistic systems, we note that
experimental results showed the existence of a \emph{circabidian}
rhythm, with a 48 h period (double the circadian period). It has been
reported that such circabidian rhythms can occasionally emerge under
controlled conditions \cite{Honma:1988:Circabidian}. The bidian rhythm
may result from a period-doubling bifurcation of a circadian rhythm
that is close to a bifurcation point.

Regarding stochasticity, gene expression is subject to two type of
noise (stochasticity): one from the discrete molecular species (intrinsic
noise) and the other from environmental variability (extrinsic noise)~\cite{Koern:2005:GeneNoiseReview,Patnaik:2006:NoiseReview,Maheshri:2007:LivingNoisyGene,Rausenberger:2009:GeneNoiseReview}.
We may take into account stochastic effects in order to make our findings
biochemically realistic. For example, a collection of identical oscillators
may exhibit damped oscillation as the averaging effect of stochasticity~\cite{Diambra:2012:DrosphilaClock}.
Nevertheless, we employed a deterministic model (i.e., ordinary differential
equations) in our analysis for two reasons. First, detailed stochastic
modeling, such as a continuous-time discrete Markov chain, is not
tractable: the only solution is an exhaustive, time-consuming calculation
using the Monte Carlo method without any theoretical guarantees. Second,
without analytical calculations, the theoretical arguments for the
enhancement would be difficult. The enhancement we detected at the
bifurcation on the limit cycle (the saddle node bifurcation) is unlikely
to be found by numerical simulations, but can be deduced from Floquet
multipliers. Although our present analyses do not incorporate stochastic
aspects, we consider that our results help in the understanding of
the stochastic dynamics in modeling mismatched periods. We will consider
such stochasticity in a future study.

Another topic yet to be investigated is when three or more oscillators
have mismatched periods. When considering more than two oscillators,
besides the oscillatory model, the topology of the coupling is important.
We can make an educated guess at the consequences of a period mismatch
in such a multioscillator model: if period inconsistency drives the
coupled system to the vicinity of bifurcation on the limit cycle,
the period mismatch will enhances entrainability. As long as the overall
connection of the multiple oscillators can be simplified as ``two
mismatched oscillators'', the entrainability of such a component
is expected to be enhanced.

In summary, we have used phase reduction and Floquet multipliers to
show that entrainability is enhanced by mismatched periods in both
smooth and relaxation oscillators. We focused on two identical oscillators
that were symmetrically coupled to each other by using a deterministic
approach. In order to make our results more useful for real biological
systems, however, it would be important to consider situations of
three or more oscillators, asymmetric coupling, or structurally heterogeneous
cases with stochasticity. These extensions are left to our future
studies.

\section*{Acknowledgments}

This work was supported by the Global COE program ``Deciphering Biosphere
from Genome Big Bang'' from the Ministry of Education, Culture, Sports,
Science and Technology (MEXT), Japan (YH and MA); a Grant-in-Aid for
Young Scientists B (\#23700263) from MEXT, Japan (YH); and a Grant-in-Aid
for Scientific Research on Innovative Areas ``Biosynthetic Machinery''
(\#11001359) from MEXT, Japan (MA).


\begin{thebibliography}{10}

\bibitem{Novak:2008:BiochemOsc}
B.~Nov{\'a}k and J.~J. Tyson.
\newblock Design principles of biochemical oscillators.
\newblock {\em Nat. Rev.}, 9:981--991, 2008.

\bibitem{Roenneberg:1993:TwoOsc}
T.~Roenneberg and D.~Morse.
\newblock Two circadian oscillators in one cell.
\newblock {\em Nature}, 362:362--364, 1993.

\bibitem{Morse:1994:TwoCircadian}
D.~Morse, J.~W. Hastings, and T.~Roenneberg.
\newblock Different phase responses of the two circadian oscillators in
  {Gonyaulax}.
\newblock {\em J. Biol. Rhythms}, 9:263--274, 1994.

\bibitem{Gonze:2000:Entrainment}
D.~Gonze and A.~Goldbeter.
\newblock Entrainment versus chaos in a model for a circadian oscillator driven
  by light-dark cycles.
\newblock {\em J. Stat. Phys.}, 101:649--663, 2000.

\bibitem{Pederse:2005:MultipleCircadian}
D.~Bell-Pedersen, V.~M. Cassone, D.~J. Earnest, S.~S. Golden, P.~E. Hardin,
  T.~L. Thomas, and M.~J. Zoran.
\newblock Circadian rhythms from multiple oscillators: lessons from diverse
  organisms.
\newblock {\em Nat. Rev.}, 6:544--556, 2005.

\bibitem{Goodwin:1965:Oscillator}
B.~C. Goodwin.
\newblock Oscillatory behavior in enzymatic control processes.
\newblock {\em Advances in Enzyme Regulation}, 3:425--437, 1965.

\bibitem{Wagner:2005:TwoGeneOsc}
A.~Wagner.
\newblock Circuit topology and the evolution of robustness in two-gene
  circadian oscillators.
\newblock {\em PNAS}, 102:11775--11780, 2005.

\bibitem{Trane:2007:RobustCircadian}
C.~Tran\'e and E.~W. Jacobsen.
\newblock On robustness as the rationale behind multiple feedback loops in the
  circadian clock.
\newblock In {\em Proceedings of Foundations of Systems Biology in
  Engineering}, 2007.

\bibitem{Hafner:2010:MultiLoop}
M.~Hafner, P.~Sacr\'e, L.~Symul, R.~Sepulchre, and H.~Koeppl.
\newblock Multiple feedback loops in circadian cycles: Robustness and
  entrainment as selection criteria.
\newblock In {\em Seventh International Workshop on Computational Systems
  Biology}, pages 43--46, 2010.

\bibitem{Hastings:2000:TwoLoopReview}
M.~H. Hastings.
\newblock Circadian clockwork: two loops are better than one.
\newblock {\em Nat. Rev.}, 1:143--146, 2000.

\bibitem{Benzi:1981:SR}
R.~Benzi, A.~Sutera, and A.~Vulpiani.
\newblock The mechanism of stochastic resonance.
\newblock {\em J. Phys. A}, 14:L453, 1981.

\bibitem{McNamara:1989:SR}
B.~McNamara and K.~Wiesenfeld.
\newblock Theory of stochastic resonance.
\newblock {\em Phys. Rev. A}, 39:4854--4869, 1989.

\bibitem{Jung:1991:AmpSR}
P.~Jung and P.~H\"anggi.
\newblock Amplification of small signals via stochastic resonance.
\newblock {\em Phys. Rev. A}, 44:8032--8042, 1991.

\bibitem{Jung:1992:GlobalSR}
P.~Jung, U.~Behn, E.~Pantazelou, and F.~Moss.
\newblock Collective response in globally coupled bistable systems.
\newblock {\em Phys. Rev. A}, 46:R1709--R1712, 1992.

\bibitem{Gammaitoni:1998:SR}
L.~Gammaitoni, P.~H\"anggi, P.~Jung, and F.~Marchesoni.
\newblock Stochastic resonance.
\newblock {\em Rev. Mod. Phys.}, 70:223--287, 1998.

\bibitem{McDonnell:2008:SRBook}
M.~D. McDonnell, N.~G. Stocks, C.~E.~M. Pearce, and D.~Abbott.
\newblock {\em Stochastic resonance}.
\newblock Cambridge University Press, 2008.

\bibitem{McDonnell:2009:SR}
M.~D. McDonnell and D.~Abbott.
\newblock What is stochastic resonance? definitions, misconceptions, debates,
  and its relevance to biology.
\newblock {\em PLoS Comput. Biol.}, 5:e1000348, 2009.

\bibitem{Hasegawa:2011:BistableSIN}
Y.~Hasegawa and M.~Arita.
\newblock Escape process and stochastic resonance under noise-intensity
  fluctuation.
\newblock {\em Phys. Lett. A}, 375:3450--3458, 2011.

\bibitem{Astumian:1994:MolecularMotor}
R.~D. Astumian and M.~Bier.
\newblock Fluctuation driven ratchets: molecular motors.
\newblock {\em Phys. Rev. Lett.}, 72:1766--1769, 1994.

\bibitem{Astumian:1997:BrownMotor}
R.~D. Astumian.
\newblock Thermodynamics and kinetics of a {Brownian} motor.
\newblock {\em Science}, 276:917--922, 1997.

\bibitem{Frey:2005:BrownianRev}
E.~Frey and K.~Kroy.
\newblock {Brownian} motion: a paradigm of soft matter and biological physics.
\newblock {\em Ann. Phys.}, 14:20--50, 2005.

\bibitem{Hanggi:2009:BrownianMotorsReview}
P.~H\"anggi and F.~Marchesoni.
\newblock Artificial {Brownian} motors: Controlling transport on the nanoscale.
\newblock {\em Rev. Mod. Phys.}, 81:387--442, 2009.

\bibitem{Hasegawa:2012:Motor}
Y.~Hasegawa and M.~Arita.
\newblock Fluctuating noise drives {Brownian} transport.
\newblock {\em J. R. Soc. Interface}, 9:3554--3563, 2012.

\bibitem{Marchesoni:1996:SpatiotempSR}
F.~Marchesoni, L.~Gammaitoni, and A.~R. Bulsara.
\newblock Spatiotemporal stochastic resonance in a $\phi^4$ model of
  kink-antikink nucleation.
\newblock {\em Phys. Rev. Lett.}, 76:2609--2612, 1996.

\bibitem{Nakao:2007:NISinLC}
H.~Nakao, K.~Arai, and Y.~Kawamura.
\newblock Noise-induced synchronization and clustering in ensembles of
  uncoupled limit-cycle oscillators.
\newblock {\em Phys. Rev. Lett.}, 98:184101, 2007.

\bibitem{Teramae:2004:NoiseIndSync}
J.~Teramae and D.~Tanaka.
\newblock Robustness of the noise-induced phase synchronization in a general
  class of limit cycle oscillators.
\newblock {\em Phys. Rev. Lett.}, 93:204103, 2004.

\bibitem{Tessone:2006:DiversityResonance}
C.~J. Tessone, C.~R. Mirasso, R.~Toral, and J.~D. Gunton.
\newblock Diversity-induced resonance.
\newblock {\em Phys. Rev. Lett.}, 97:194101, 2006.

\bibitem{Sharma:2003:CircadianAdaptive}
V.~K. Sharma.
\newblock Adaptive significance of circadian clocks.
\newblock {\em Chronobiol. Int.}, 20:901--919, 2003.

\bibitem{Beaver:2002:FlyClock}
L.~M. Beaver, B.~O. Gvakharia, T.~S. Vollintine, D.~M. Hege, R.~Stanewsky, and
  J.~M. Giebultowicz.
\newblock Loss of circadian clock function decreases reproductive fitness in
  males of {{\em Drosophila}} {\em melanogaster}.
\newblock {\em PNAS}, 19:2134--2139, 2002.

\bibitem{Dodd:2005:PlantCircadian}
A.~N. Dodd, N.~Salathia, A.~Hall, E.~K\'evei, R.~T\'oth, F.~Nagy, J.~M.
  Hibberd, A.~J. Millar, and A.~A.~R. Webb.
\newblock Plant circadian clocks increase photosynthesis, growth, survival and
  competitive advantage.
\newblock {\em Science}, 309:630--633, 2005.

\bibitem{Woelfle:2004:CyanoClock}
M.~A. Woelfle, Y.~Ouyang, K.~Phanvijhitsiri, and C.~H. Johnson.
\newblock The adaptive value of circadian clocks: an experimental assessment in
  cyanobacteria.
\newblock {\em Curr. Biol.}, 14:1481--1486, 2004.

\bibitem{Gonze:2005:CircadianSync}
D.~Gonze, S.~Bernard, C.~Waltermann, A.~Kramer, and H.~Herzel.
\newblock Spontaneous synchronization of coupled circadian oscillators.
\newblock {\em Biophys. J.}, 89:120--129, 2005.

\bibitem{Liu:2007:Cell}
A.~C. Liu, D.~K. Welsh, C.~H. Ko, H.~G. Tran, E.~E. Zhang, A.~A. Priest, E.~D.
  Buhr, O.~Singer, K.~Meeker, I.~M. Verma, F.~J. {Doyle III}, J.~S. Takahashi,
  and S.~A. Kay.
\newblock Intercellular coupling confers robustness against mutations in the
  {SCN} circadian clock network.
\newblock {\em Cell}, 129:605--616, 2007.

\bibitem{Tony:2008:TunablePositive}
T.~Y.-C. Tsai, Y.~S. Choi, W.~Ma, J.~R. Pomerening, C.~Tang, and J.~E. {Ferrell
  Jr.}
\newblock Robust, tunable biological osillations from interlinked positive and
  negative feedback loops.
\newblock {\em Science}, 321:126--129, 2008.

\bibitem{Zhou:2008:SyncGeneOsc}
T.~Zhou, J.~Zhang, Z.~Yuan, and L.~Chen.
\newblock Synchoronization of genetic oscillators.
\newblock {\em Chaos}, 18:037126, 2008.

\bibitem{Komin:2010:EntrainCircadian}
N.~Komin, A.~C. Murza, E.~Hern\'andez-Garc\'ia, and R.~Toral.
\newblock Synchronization and entrainment of coupled circadian oscillators.
\newblock {\em Interface Focus}, 1:167--176, 2010.

\bibitem{Elowitz:2000:Repressilator}
M.~B. Elowitz and S.~Leibler.
\newblock A synthetic oscillatory network of transcriptional regulators.
\newblock {\em Nature}, 403:335--338, 2000.

\bibitem{FitzHugh:1961:FHmodel}
R.~FitzHugh.
\newblock Impulses and physiological states in theoretical models of nerve
  membrane.
\newblock {\em Biophys. J.}, 1:445--466, 1961.

\bibitem{Nagumo:1962:NagumoModel}
J.~Nagumo, S.~Arimoto, and S.~Yoshizawa.
\newblock An active pulse transmission line simulating nerve axon.
\newblock {\em Proc. IRE}, 50:2061--2070, 1962.

\bibitem{Kuramoto:2003:OscBook}
Y.~Kuramoto.
\newblock {\em Chemical Oscillations, Waves, and Turbulence}.
\newblock Dover publications, Mineola, New York, 2003.

\bibitem{Izhikevich:2007:NeuroBook}
E.~M. Izhikevich.
\newblock {\em Dynamical Systems in Neuroscience: The Geometry of Excitability
  and Bursting}.
\newblock MIT Press, 2007.

\bibitem{Pikovsky:2001:SyncBook}
A.~Pikovsky, M.~Rosenblem, and J.~Kurths.
\newblock {\em Synchronization: A universal concept in nonlinear sciences}.
\newblock Cambridge University Press, 2001.

\bibitem{Shwemmer:2012:WeaklyCoupleOsc}
M.~A. Schwemmer and T.~Lewis.
\newblock The theory of weakly coupled oscillators.
\newblock In N.~Schultheiss, A.~Prinz, and R.~Butera, editors, {\em Phase
  response curves in Neuroscience: Theory, Experiment and Analysis}, pages
  3--31. Springer, 2012.

\bibitem{Klausmeier:2008:Floquet}
C.~A. Klausmeier.
\newblock Floquet theory: a useful tool for understanding nonequilibrium
  dynamics.
\newblock {\em Theor. Ecol.}, 1:153--161, 2008.

\bibitem{Guckenheimer:1983:DynsysBook}
J.~Guckenheimer and P.~Holmes.
\newblock {\em Nonlinear Oscillations, Dynamical Systems, and Bifurcations of
  Vector Fields}.
\newblock Springer, 1983.

\bibitem{Crawford:1991:IntroBifTheory}
J.~D. Crawford.
\newblock Introduction to bifurcation theory.
\newblock {\em Rev. Mod. Phys.}, 63:991--1037, 1991.

\bibitem{Wiesenfeld:1985:PDBifAmp}
K.~Wiesenfeld and B.~McNamara.
\newblock Period-doubling systems as small-signal amplifiers.
\newblock {\em Phys. Rev. Lett.}, 55:13--16, 1985.

\bibitem{Wiesenfeld:1986:SignalAmpBif}
K.~Wiesenfeld and B.~McNamara.
\newblock Small-signal amplification in bifurcating dynamical systems.
\newblock {\em Phys. Rev. A}, 33:629--642, 1986.

\bibitem{Honma:1988:Circabidian}
K.~Honma and S.~Honma.
\newblock Circabidian rhythm: its appearance and disappearance in association
  with a bright light pulse.
\newblock {\em Experientia}, 44:981--983, 1988.

\bibitem{Koern:2005:GeneNoiseReview}
M.~K{\oe}rn, T.~C. Elston, W.~J. Blake, and J.~J. Collins.
\newblock Stochasticity in gene expression: from theories to phenotypes.
\newblock {\em Nat. Rev.}, 6:451--464, 2005.

\bibitem{Patnaik:2006:NoiseReview}
P.~R. Patnaik.
\newblock External, extrinsic and intrinsic noise in cellular systems:
  analogies and implications for protein synthesis.
\newblock {\em Biotechnol. Mol. Biol. Rev.}, 1:121--127, 2006.

\bibitem{Maheshri:2007:LivingNoisyGene}
N.~Maheshri and E.~K. O'Shea.
\newblock Living with noisy genes: How cells function reliably with inherent
  variability in gene expression.
\newblock {\em Annu. Rev. Biophys. Biomol. Struct.}, 36:413--434, 2007.

\bibitem{Rausenberger:2009:GeneNoiseReview}
J.~Rausenberger, C.~Fleck, J.~Timmer, and M.~Kollmann.
\newblock Signatures of gene expression noise in cellular systems.
\newblock {\em Prog. Biophys. Mol. Biol.}, 100:57--66, 2009.

\bibitem{Diambra:2012:DrosphilaClock}
L.~Diambra and C.~P. Malta.
\newblock Modeling the emergence of circadian rhythms in a clock neuron
  network.
\newblock {\em PLoS One}, 7:e33912, 2012.

\end{thebibliography}
\end{document}